\documentclass
[aps,pra,superscriptaddress,floatfix,showpacs,10pt,twocolumn]{revtex4}%
\usepackage{amsmath}
\usepackage{graphicx}
\usepackage{amsfonts}
\usepackage{amssymb}
\usepackage[T1]{fontenc}
\usepackage[latin2]{inputenc}
\usepackage{latexsym}
\usepackage{bbm}
\usepackage{color}%
\begin{document}
\title{Motion-enhanced quantum entanglement in the dynamics of excitation
transfer}
\author{Wei Song}
\affiliation{Institute for Quantum Control and Quantum Information, and School of Electronic and Information Engineering, Hefei Normal University, Hefei 230601, China}
\author{Yi-Sheng Huang}
\affiliation{School of Physics and Material Science, Anhui University, Hefei 230601, China }
\author{Ming Yang}
\affiliation{School of Physics and Material Science, Anhui University, Hefei 230601, China }
\author{Zhuo-Liang Cao}
\affiliation{Institute for Quantum Control and Quantum Information, and School of Electronic and Information Engineering, Hefei Normal University, Hefei 230601, China}

\date{\today}

\begin{abstract}
We investigate the dynamics of entanglement in the excitation
transfer through a model consisting of three interacting molecules coupled
to environments. It is shown that entanglement can be further enhanced if the distance between the molecules is oscillating. Our results demonstrate that motional effect plays a constructive role on quantum entanglement in the dynamics of excitation
transfer. This mechanism might provide useful guideline for designing artificial systems to battle against decoherence.
\end{abstract}
\maketitle

Usually quantum entanglement is regarded as a fragile resource and very sensitive to noise. Thus we need rigorous laboratory conditions to manufacture and maintain entanglement. It is believed that entanglement cannot exist outside the
laboratories, not to mention biological systems, which are wet and hot, and with extremely high levels of noise\cite{Briegel:2008}. However, several recent studies have shown that entanglement might exist in non-equilibrium systems and survive for a relatively long time scales at physiological temperatures in biological systems. Up to date, these studies can be divided into three aspects:  the first aspect is focused on quantum entanglement in model systems. The first study was made by M. Thorwart \emph{et al}\cite{Thorwart:2009} who demonstrate that enhanced quantum entanglement in the non-Markovian dynamics of biomolecular excitons. It is shown in Ref. \cite{Cai:2010} that dynamic entanglement can be continuously generated in noisy non-equilibrium systems. Subsequently, this phenomenon has been generalized to the spin gas model and non-Markovian models \cite{Guerreschi:2012}. F. Galve \emph{et al} \cite{Galve:2010} also predict that
nanomechanical oscillators can be entangled
at much higher temperatures than previously
thought possible.

The second aspect is quantum entanglement in Fenna-Matthews-Olson (FMO) complex of green sulfur bacterium. The FMO complex is a water soluble complex and acts as a molecular wire to transfer the excitation energy from the light harvesting antenna to the reaction center. Many theoretical studies in light harvesting structures are focused on the FMO due to its well characterized pigment-protein structure. Recent experiments by G. S. Engel \emph{et al} \cite{Engel:2007} and G. Panitchayangkoon \emph{et al} \cite{Panitchayangkoon:2010} have shown that the electronic coherence between two excitonic levels at both cryogenic (77K)
and ambient (300K) temperatures. These experimental studies have generated a lot of theoretical interests\cite{Plenio:2008,Nalbach:2011,Ai:2012,Castro:2008,Kassal:2013,Yang:2010,Chin:2010,Liang:2010,Ishizakia:2009,Ghosh:2011,Tan:2012,Rey:2013,Asadian:2010,Li:2013,Qin:2014,Zhang:2013,Wang:2013} in understanding the role of quantum coherence in light harvesting efficiency. In particular, M. Sarovar \emph{et al} \cite{Sarovar:2010} present numerical evidences for the existence of entanglement in the FMO complex for relatively long times. The influence of Markovian, as well as non-Markovian noise on the dynamics of entanglement in FMO complex have also been analyzed
by F. Caruso \emph{et al.} \cite{Caruso:2010}.

The third aspect includes the investigation of entanglement in larger light harvesting complexes. Light harvesting complex II (LHCII) is the most abundant photosynthetic antenna complex in plants containing over 50$\%$ of the world's chlorophyll
molecules \cite{Amerongen:2000,Blankenship:2002}. It is shown that LHCII also exhibits long-lived electronic coherence. Ishizaki and Fleming  \cite{Ishizaki:2010}have investigated quantum entanglement in LHCII across different bipartitions of the chlorophyll pigments.

However, in previous investigations listed above, the motional effect of the molecules in the dynamics of excitation
transfer is omitted. It is suggested that conformational motion is an important
feature of molecular processes, such as protein folding or excitation
transfer in light harvesting complex. These lead to effectively time-dependent interactions, with their strengths modulated by the motion of the molecular. In analogy to the static case, the effect of the motion on the entanglement dynamics of excitation transfer is far from being understood. Here, we tackle this problem by considering a model consisting of three interacting molecules driven through the oscillating motion. Our studies demonstrate that for a wide range of parameters, the average entanglement can be enhanced if the motional effect is taken into account. In our discussion, we assume that the conformational motion of the molecular structure can be described classically. This semi-quantal approximation holds if the involved molecules are too large to show their quantum behavior\cite{Guerreschi:2012}. Motion of the individual molecules leads to a change in dipole moments and thus also induces a time-dependence of the coupling. This mechanism will be dominant whenever the molecules
are tightly embedded in a protein scaffold such as in the FMO complex. For simplicity, we only consider the modulation of
coupling strength due to the change in distance. According to Ref.\cite{Guerreschi:2012}, we choose the motion
of the molecules change their distance periodically, which holds for small
amplitudes in the harmonic regime. Furthermore, we suppose the molecules are distant enough from each other such that we only consider the nearest neighbor interactions. In natural conditions, it is reasonable to consider at most one excitation during the excitation transfer process\cite{Sarovar:2010}.

Under these conditions, the Hamiltonian of the chain of interacting molecules in the single-excitation manifold can be written as

\begin{align}
H = \sum\limits_{n = 1}^N {\varepsilon _n } \left| n \right\rangle \left\langle n \right| + \sum\limits_{n = 1}^{N - 1} {J_n } \left( {\left| n \right\rangle \left\langle {n + 1} \right| + \left| {n + 1} \right\rangle \left\langle n \right|} \right)
\end{align}

\noindent where $\left| n \right\rangle$ represents the state with the excitation at the $n$-th site having energy ${\varepsilon _n }$  and all other states are in their electronic ground state, and ${J_n }$
is the coupling strength between the $n$-th and the $(n+1)$-th molecule. Motion of the molecules will induce a time-dependent coupling strength and deformation of the molecules will also lead to the change of dipolar coupling. For simplicity, we suppose the coupling strength only changes with the relative distance between the molecules. Suppose the distance between the molecules $n$ and $n+1$ is $d_n \left( t \right) = d_0  - \left[ {u_n \left( t \right) - u_{n + 1} \left( t \right)} \right] = d_0 \left[ {1 - 2a_n \sin \left( {\omega t + \phi _n } \right)} \right]$, where ${u_n \left( t \right)}$
denotes the position of the $n$-th molecule and $d_0$ is the equilibrium distance between two neighboring molecules, and $a_n$ is the individual sites' relative amplitude of
oscillation. Correspondingly, the dipole-dipole coupling strength between two molecules is given by\cite{Cai:2010,Guerreschi:2012}

\begin{align}
J_n \left( t \right) = \frac{{\tilde J_0 }}{{\left[ {d_n \left( t \right)} \right]^3 }} = \frac{{J_0 }}{{\left[ {1 - 2a_n \sin \left( {\omega t + \phi _n } \right)} \right]^3 }}
\end{align}

\noindent where we have defined $J_0  = \frac{{\tilde J_0 }}{{d_0^3 }}$, and $\tilde J_0$ contains the dipole moments and physical constants. Here our discussion corresponds
to a given amplitude, frequency, and phase synchronized with the propagation of the excitation. This assumption is reasonable
because the wave packet that describes the nuclear motion has been observed to exhibit surprisingly long coherence times in
reaction center proteins\cite{Vos:1993}. Although this model is rather simple, several biological
systems show similar structure that agree with the assumptions underlying our model. For example, a type of secondary structure
in proteins named $\alpha$-helix or Fenna-Matthews-Olson (FMO) complex, among which an excitation can be exchanged due to
dipole-dipole couplings between the molecules. Firstly, we assume that all sites are subjected to dissipative noise. This process can be introduced by considering a Lindblad term $L_{diss} \left( \rho  \right)$ of the form

\begin{align}
L_{diss} \left( \rho  \right) = \sum\limits_{n = 1}^N {\Gamma _n } \left[ { - \left\{ {\sigma _n^ +  \sigma _n^ -  ,\rho } \right\} + 2\sigma _n^ -  \rho \sigma _n^ +  } \right]
\end{align}

\noindent where $
{\sigma _n^ +   = \left| n \right\rangle \left\langle 0 \right|}$ and $
{\sigma _n^ -   = \left| 0 \right\rangle \left\langle n \right|}$ are, respectively, the raising and lowering operators
for site $n$, and ${\left| 0 \right\rangle }$ refers to the zero exciton state of the system. The symbol $\left\{ {A,B} \right\}$ is an anticommutator, and $\Gamma _n$ denotes the dissipation rate of the $n$-th molecule.

\begin{figure}[ptb]
\includegraphics[scale=0.7,angle=0]{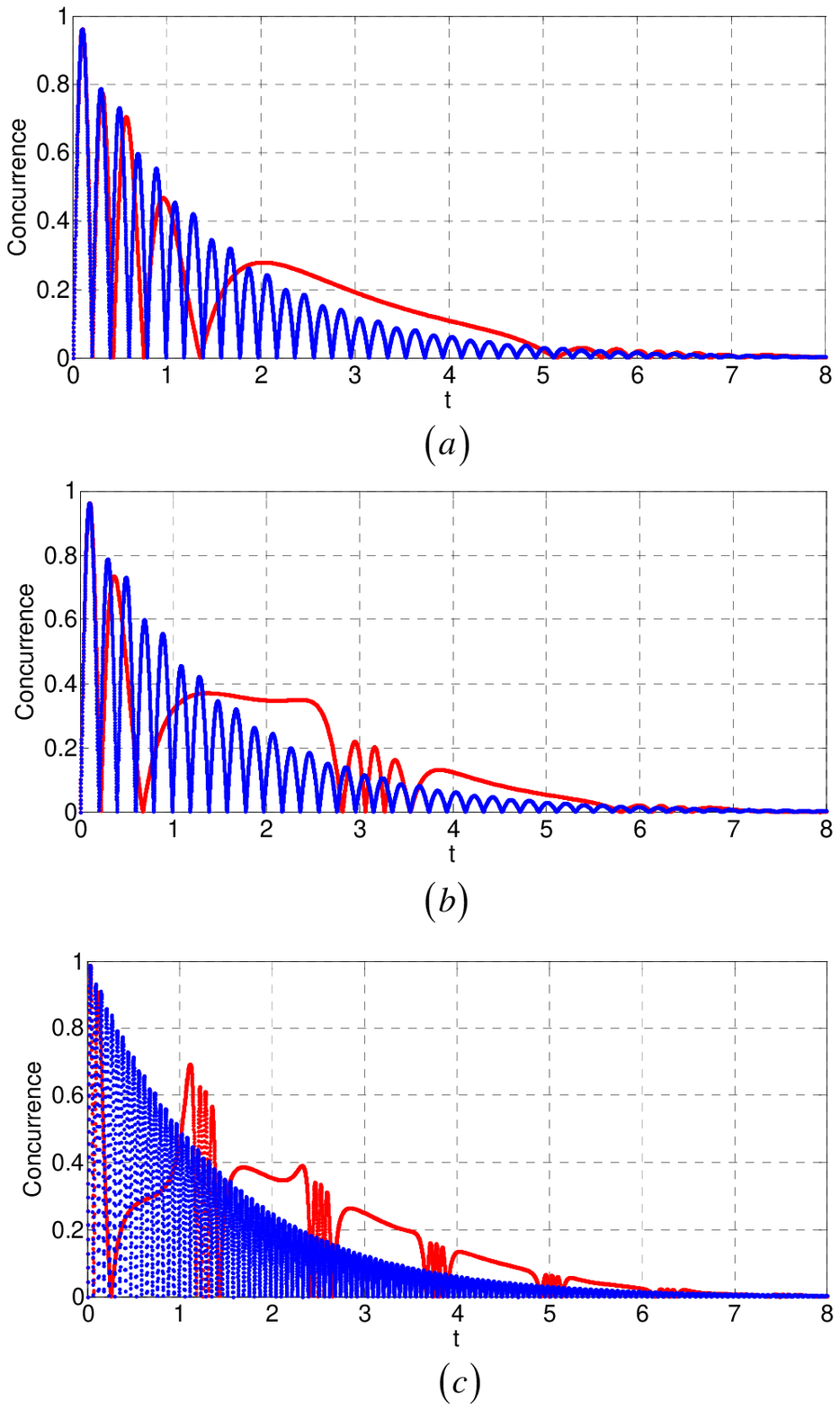}\caption{(color online). Time evolutions of the entanglement of molecules 1 and 2, in the motional case(red line), and in static case(blue line).(a)$\omega  = 1,a_1  = \frac{1}{4}$,(b)$\omega  = 2,a_1  = \frac{1}{4}$,(c)$\omega  = 5,a_1  = \frac{1}{3}$.}%
\end{figure}

In order to model the dynamics of the excitation transfer along a chain of molecules, we introduce an additional site, the sink that resembles the reaction center in photosynthesis. It denotes an irreversible decay of
excitations from the site $N$ to the last $N+1$ site. The absorption of
the energy from the site $N$ to the sink (numbered $N+1$) is modeled by a Lindblad operator

\begin{align}
L_{sink} \left( \rho  \right) = \Gamma _s \left[ {2\sigma _{N + 1}^ +  \sigma _N^ -  \rho \sigma _N^ +  \sigma _{N + 1}^ -   - \left\{ {\sigma _N^ +  \sigma _{N + 1}^ -  \sigma _{N + 1}^ +  \sigma _N^ -  ,\rho } \right\}} \right]
\end{align}

\noindent where $\Gamma _s$ is the absorption rate of the sink which describes the
irreversible decay of the excitations to the sink. Thus the master equation of the density matrix $\rho$ of the system is given by

\begin{align}
\frac{{d\rho }}{{dt}} = i\left[ {\rho ,H} \right] + L_{diss} \left( \rho  \right) + L_{sink} \left( \rho  \right)
\end{align}

To study the role of the motional effect on entanglement, we only considering the simplest case of a chain composed of only $N = 2$ interacting molecules, labeled 1 and 2 plus the sink labeled 3. The exciton is transferred from molecule 1 to 3 through the linear chain and finally is trapped by the sink. Molecules 1 and 2 are subjected to the dissipative environment simultaneously. The coupling strength between the two sites is modulated by the oscillating motion of the molecules. In our discussion, we suppose the local energies $\varepsilon _1  = \varepsilon _2  = \varepsilon$. This simple model provides a platform to demonstrate the dynamics of entanglement in the excitation transfer process driven through the
oscillating motion. Here, we use the Wootter's concurrence \cite{Wootters:1998}
to quantify entanglement, which can be defined as: $C\left( \rho\right)  = \max\left\{
{0,\sqrt{\lambda_{1} } - \sqrt{\lambda_{2} } - \sqrt{\lambda_{3} } -
\sqrt{\lambda_{4} } } \right\} $, where ${\lambda_{i} }$ are the
eigenvalues in decreasing order of the matrix $ \rho\tilde\rho=
\rho\sigma_{y} \otimes\sigma_{y} \rho^{*} \sigma_{y}
\otimes\sigma_{y}$ with $\rho^{*}$ denoting the complex conjugation
of $\rho$. We choose the parameter $\Gamma _1  = \Gamma _2  = 0.2,\Gamma _s  = 0.5, \phi  = \frac{\pi }{2},J_0  = 1$, where $\phi  = \frac{\pi }{2}$ indicates the two molecules are closest at the initial time.

We suppose the initial state has an excitation localized on molecule 1. By numerically solving the master equation (5), we plot the entanglement evolution of molecules 1 and 2 in Fig.1 with red line. In this case, the coupling strength $J\left( t \right)$
is time-dependent and driven through the oscillating motion of the molecules. In Fig.1(a) we choose the parameters $\omega  = 1, a_1  = \frac{1}{4}$ for the motional case. For comparison, we also consider the static case with constant $J$, and the coupling strength $J_{\max }  = 8$ when the molecules have the closest distance. The blue line represents the evolution of entanglement in the static case. In Fig.1 (b) and (c), we choose the parameters $\omega  = 2, a_1  = \frac{1}{4}, J_{\max }  = 8$ and $\omega  = 5,a_1  = \frac{1}{3}, J_{\max }  = 27$, respectively. Fig.1 shows that in the motional case entanglement has larger value than the static case for a wide range of time.

\begin{figure}[ptb]
\includegraphics[scale=0.7,angle=0]{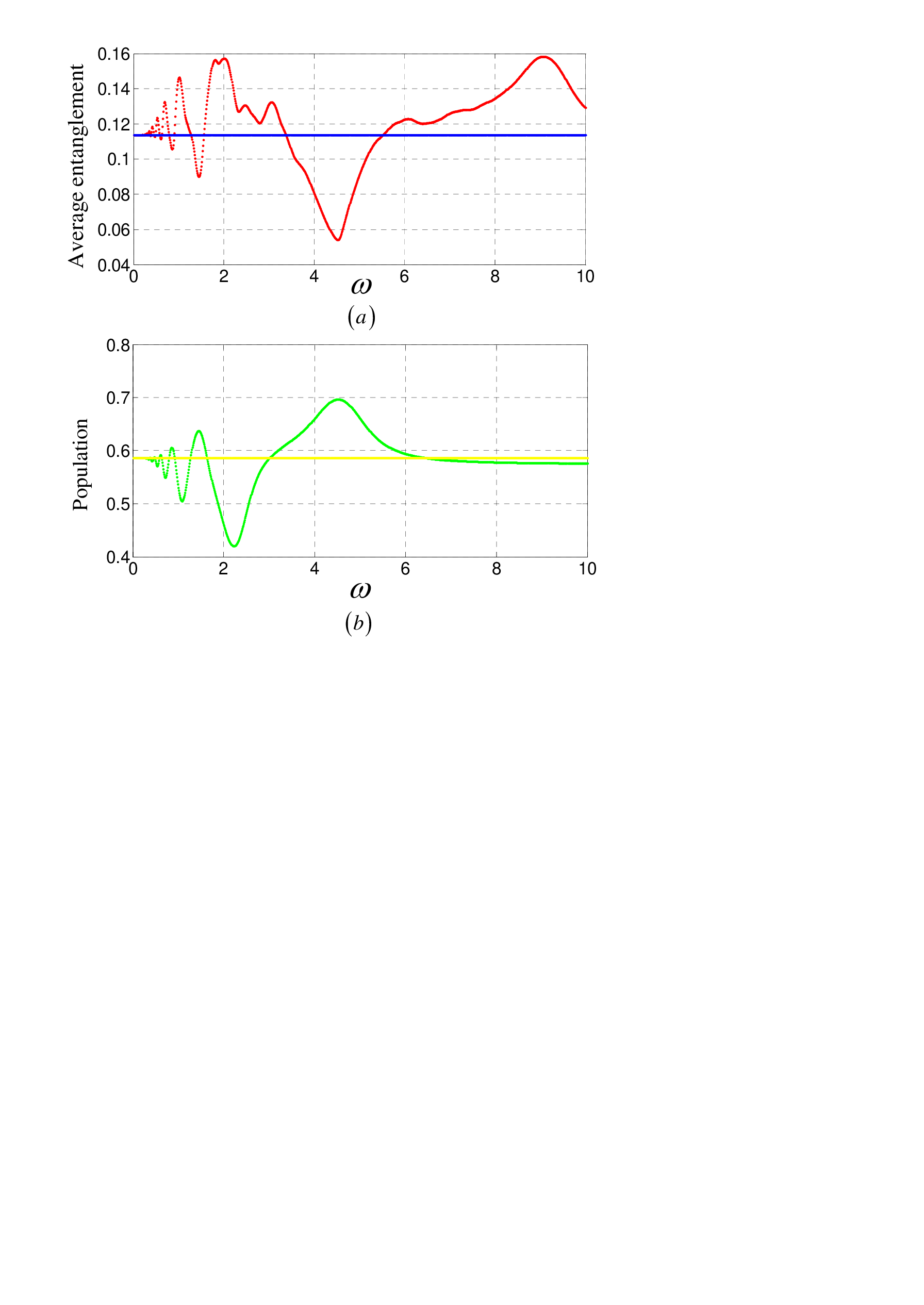}\caption{(color online). (a) The average entanglement evolutions versus $\omega$, in the motional case(red line), and in static case(blue line). (b) The average sink population evolutions versus $\omega$, in the motional case(green line), and in static case(yellow line). The parameters are $\phi  = \frac{\pi }{2}$, $a_1  = \frac{1}{4},0 \le T \le 8$.}%
\end{figure}

\begin{figure}[ptb]
\includegraphics[scale=0.8,angle=0]{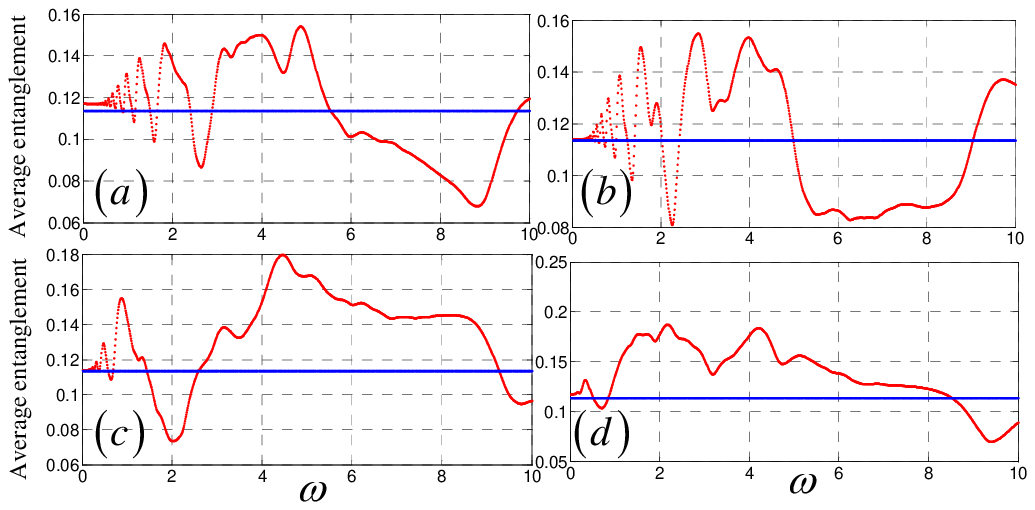}\caption{(color online). The average entanglement evolutions versus $\omega$ with different $\phi$. (a)$\phi  = \frac{\pi }{6}$, (b)$\phi  = \frac{\pi }{3 }$, (c)$\phi  = \frac{{2\pi }}{3}$, (d)$\phi  = \frac{{5\pi }}{6}$.}%
\end{figure}

In order to further illustrate the dynamics of entanglement with different $\omega$, we introduce the average entanglement defined as $\bar C = \frac{1}{T}\int_0^T {C\left( t \right)dt}$ for fixed $\omega$. From Fig.1(a) we can calculate that $\bar C\left( {J_0 } \right) = {\rm{0}}{\rm{.1461 > }}\bar C\left( {J_{\max } } \right) = {\rm{0}}{\rm{.1136}}$, where we have chosen the time in the range ${\rm{0}} \le {\rm{T}} \le {\rm{8}}$. Similarly, in cases (b) and (c) we have $\bar C\left( {J_0 } \right) = {\rm{0}}{\rm{.1571 > }}\bar C\left( {J_{\max } } \right) = {\rm{0}}{\rm{.1136}}$ and $\bar C\left( {J_0 } \right) = {\rm{0}}{\rm{.1637 > }}\bar C\left( {J_{\max } } \right) = {\rm{0}}{\rm{.1133}}$, respectively, with numerical calculations. Obviously, the average entanglement is larger than the static case in these three cases. In Fig.2 (a) we plot the average entanglement versus frequency $\omega$ with ${\rm{0}} \le {\omega} \le {\rm{10}}$. The red line corresponds to the time-dependent coupling strength and the blue line represents the static case. We can see that the average entanglement oscillate rapidly around the static case for small $\omega$. For fast motion, i.e. for larger $\omega$, the oscillating motion plays a constructive role in contrast to the static case. This enhancement is valid only for a window of $\omega$. This effect can be understood as follows. The oscillating of the molecule leads to a population redistribution, as compared to the static case. The enhancement of entanglement corresponds to a relatively larger population in the molecule 1 and 2. It means that motion might prevent excitation transfer from the molecule chain to the sink for some $\omega$. Thus the enhancement of entanglement is accompanied with the decrease of the sink population. This explanation can be confirmed in Fig.2 (b), in which we plot the average evolution of sink population versus $\omega$. Here, the sink population at time $t$ is defined as $p\left( t \right) = Tr\left( {\left| 3 \right\rangle \left\langle 3 \right|\rho \left( t \right)} \right)$, where ${\left| 3 \right\rangle }$ denotes the sink. The average sink population is given by $\bar p = \frac{1}{T}\int_0^T {p\left( t \right)} dt$. It can be seen from Fig.2 (b) that entanglement is enhanced when the population is decreased with fixed $\omega$. There exist a tradeoff relation between the average entanglement and the average sink population.

The previous discussion was limited to the initial condition $\phi  = \frac{\pi }{2}$ for the molecules in the closest distance. In order to test whether our results hold for other phases, we set $\phi  = \frac{\pi }{6}$, $\phi  = \frac{\pi }{3}$, $\phi  = \frac{{2\pi }}{3}$, and $\phi  = \frac{{5\pi }}{6}$, respectively in Fig.3. It shows that the enhancement of entanglement still exist for other phases. If $0 \le \phi  \le \pi$, the evolution of entanglement strongly modulated by the initial phases and the window of $\omega$ increases with the initial phase. As seen in Fig.3 (d) that entanglement is larger than the static case for almost all $\omega$. In Fig.4 we also plot the evolution of average entanglement with different $a_1$, the other parameters are $\Gamma _1  = \Gamma _2  = 0.2,\Gamma _s  = 0.5, \phi  = \frac{\pi }{2},J_0  = 1,\omega  = 1$. Here, the maximal coupling strength in the static case is given by $J_{\max }  = \frac{{J_0 }}{{\left( {1 - 2a_1 } \right)^3 }}$.

Finally, we investigate whether our results also hold for other initial states. Firstly, the initial state is set in the maximal superposition state $
\frac{1}{{\sqrt 2 }}\left( {\left| 1 \right\rangle  + \left| 2 \right\rangle } \right)$, the other parameters are $\phi  = \frac{\pi }{2}$, $a_1  = \frac{1}{4},0 \le T \le 8$. Fig. 5 (a) and (b) shows the evolutions of entanglement versus $\omega$ with $\phi  = \frac{\pi }{6}$, and $\phi  = \frac{{5\pi }}{6}$, respectively. Obviously, the average entanglement is always larger than the static case for this initial state. The parameters in Fig.5 (c) and (d) are similar to (a) and (b) except the initial state is ${\sqrt {\frac{1}{3}} \left| 1 \right\rangle  + \sqrt {\frac{2}{3}} \left| 2 \right\rangle }$. Fig.5 shows that when the initial state is a superposition state, motion helps to enhance entanglement better than the inital state ${\left| 1 \right\rangle }$.

In the above discussions, we have modeled the environment as dissipative noise. In order to test how the dephasing process affects the dynamics of entanglement, we can add an additional dephasing term in Eq.(5) which is given by $L_{deph} \left( \rho  \right) = \sum\limits_{n = 1}^N {\gamma _n \left( {2\sigma _n^ +  \sigma _n^ -  \rho \sigma _n^ +  \sigma _n^ -   - \left\{ {\sigma _n^ +  \sigma _n^ -  ,\rho } \right\}} \right)}$. Numerical simulation shows that the dephasing noise cannot alter our results and dephasing only affects the amplitude of entanglement.

In summary, we have investigated the dynamics of entanglement in the excitation
transfer through a model consisting of three interacting molecules coupled
to environments. The results presented here
demonstrate that quantum entanglement can be enhanced through the mechanical motion of the molecules.
Our model can be simulated with trapped ions and the two internal levels of trapped ions can encode the two-level system. The classical oscillation can be
simulated by tuning the interaction strength and the transverse fields, which is achievable, e.g., by changing the amplitudes of
laser beams as suggested in Ref.\cite{Cai:2010}. In biologically context, such mechanical motion can be regarded as conformational changes which plays a vital role in many molecular processes. Here, we model these conformational changes as purely classical oscillating of the molecules. This effect suggests us that biological systems might utilize this mechanics to protect entanglement in natural environments. Our results may have potential applications in future artificial systems to maintain entanglement in the presence of environment noises.

\begin{figure}[ptb]
\includegraphics[scale=0.75,angle=0]{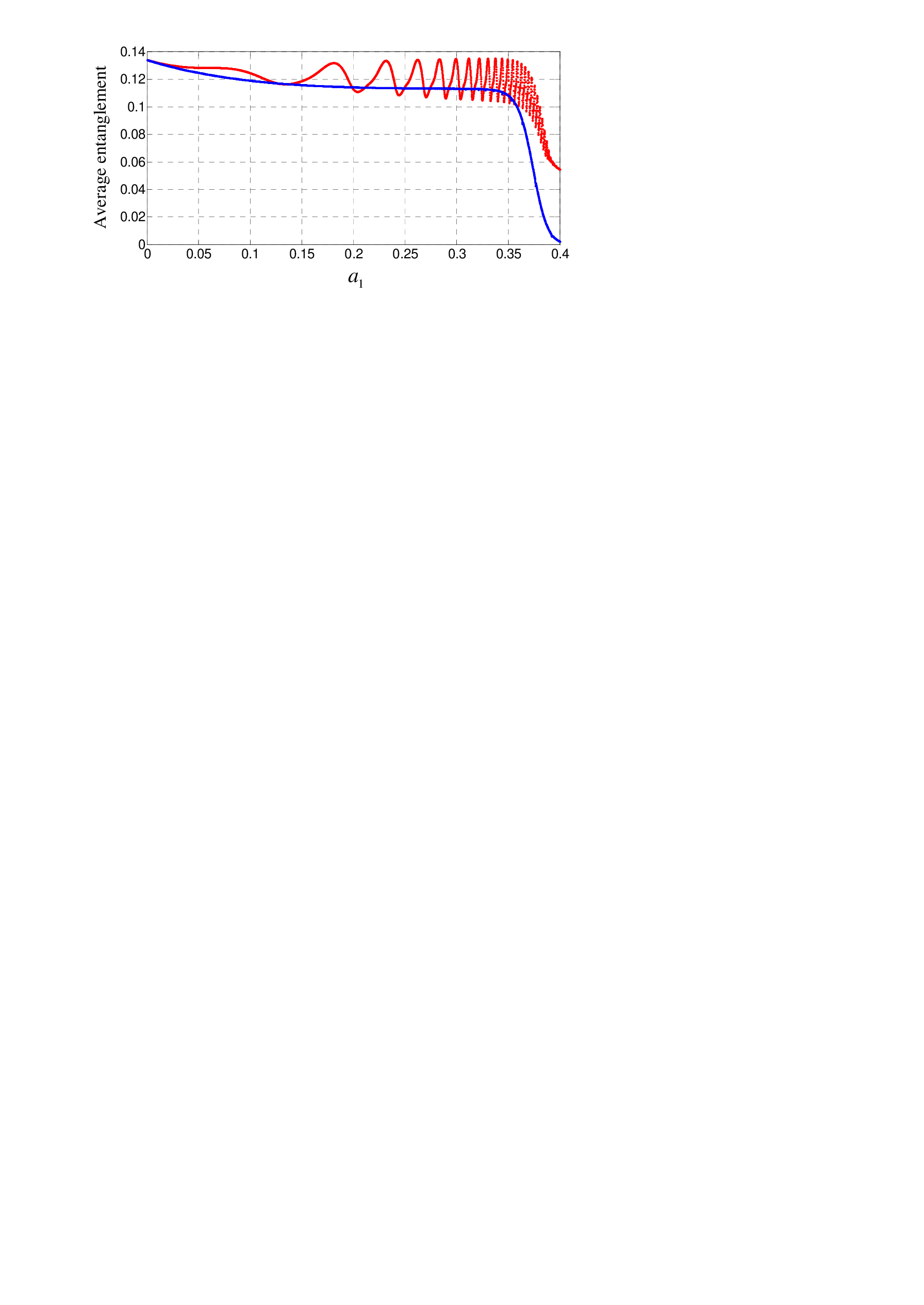}\caption{(color online). The average entanglement evolutions versus $a_1$ with $\phi  = \frac{\pi }{2}, \omega  = 1$.}%
\end{figure}
\begin{figure}[ptb]
\includegraphics[scale=0.75,angle=0]{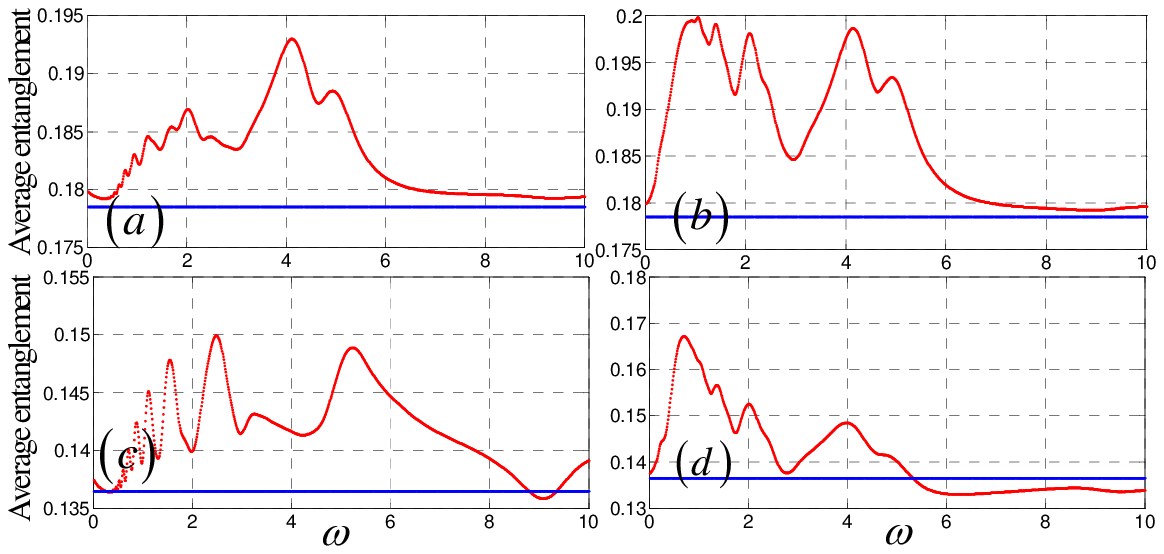}\caption{(color online). The average entanglement evolutions versus $\omega$ for the initial state $\frac{1}{{\sqrt 2 }}\left( {\left| 1 \right\rangle  + \left| 2 \right\rangle } \right)$:(a)$\phi  = \frac{\pi }{6}$,(b)$\phi  = \frac{{5\pi }}{6}$. The parameters in (c) and (d) are similar to (a) and (b) except the initial state is $
{\sqrt {\frac{1}{3}} \left| 1 \right\rangle  + \sqrt {\frac{2}{3}} \left| 2 \right\rangle }$.}%
\end{figure}

We thank Xiao-Li Huang for helpful discussions on numerical simulations. This work was supported by National Natural Science Foundation of
China under Grant No.11374085, No.61073048, No.11274010, the Specialized Research Fund for the
Doctoral Program of Higher Education(no. 20113401110002), the '211'
Project of Anhui University, the Anhui Provincial Natural Science Foundation 1408085MA20, the personnel department of Anhui
province and the 136 Foundation of Hefei Normal University under
Grant No.2014136KJB04.


\begin{thebibliography}{99}                                                                                               %
\bibitem {Briegel:2008}Briegel H J and Popescu S 2008 arXiv:0806.4552[quant-ph].

\bibitem {Thorwart:2009}Thorwart M, Eckel J, Reina J H, Nalbach P and Weiss S 2009 Chem. Phys. Lett. \textbf{478} 234.

\bibitem {Cai:2010}Cai Jianming, Popescu S, Briegel H J 2010 Phys. Rev. E \textbf{82} 021921.

\bibitem {Guerreschi:2012}Guerreschi G G, Cai Jianming, Popescu S and Briegel H J 2012 New J. Phys. \textbf{14} 053043.

\bibitem {Galve:2010}Galve F, Pachon L A and Zueco D 2010 Phys. Rev. Lett. \textbf{105} 180501.

\bibitem {Engel:2007}Engel G S, Calhoun T R, Read E L, Ahn T K, Mancal T, Cheng Y C, Blankenship R E and Fleming G R Nature. 2007 \textbf{446} 782.

\bibitem {Panitchayangkoon:2010}Panitchayangkoon G, Hayes D, Fransted K A, Caram J R, Harel E, Wen J, Blankenship R E, Engel G S 2010 Proc. Natl. Acad. Sci. USA. \textbf{107} 12766.

\bibitem {Plenio:2008}Plenio M B and Huelga S F 2008 New J. Phys. \textbf{10} 113019.

\bibitem {Nalbach:2011}Nalbach P, Braun D, Thorwart M 2011 Phys. Rev. E \textbf{84} 041926.

\bibitem {Ai:2012}Ai Bao-quan and Zhu Shi-Liang 2012 Phys. Rev. E \textbf{86} 061917.

\bibitem {Castro:2008}Castro A Olaya, Lee C F, Olsen F F and Johnson N F 2008 Phys. Rev. B \textbf{78}, 085115.

\bibitem {Kassal:2013}Kassal I, Zhou J Y and Keshari S R 2013 J. Phys. Chem. Lett. \textbf{4} 362.

\bibitem {Yang:2010}Yang S, Xu D Z, Song Z and Sun C P 2010 J. Chem. Phys. \textbf{132} 234501.

\bibitem {Chin:2010}Chin A W, Datta A, Caruso F, Huelga S F and Plenio M B 2010 New J. Phys. \textbf{12} 065002.

\bibitem {Liang:2010}Liang Xian-Ting 2010 Phys. Rev. E \textbf{82} 051918.


\bibitem {Ishizakia:2009}Ishizakia A and Fleming G R 2009 Proc. Natl. Acad. Sci. USA. \textbf{106} 17255.

\bibitem {Ghosh:2011}Ghosh P K, Smirnov A Y and Nori F 2011 J. Chem. Phys. \textbf{134} 244103.


\bibitem {Tan:2012}Tan Qing-Shou and Kuang Le-Man 2012 Commun. Theor. Phys. \textbf{58} 359.


\bibitem {Rey:2013}Rey M, Chin A W, Huelga S F and Plenio M B 2013 J. Phys. Chem. Lett. \textbf{4} 903.


\bibitem {Asadian:2010}Asadian A, Tiersch M, Guerreschi G G, Cai Jianming, Popescu S and Briegel H J 2010 New J. Phys. \textbf{12} 075019.

\bibitem {Li:2013}Li H R, Zhang Pei, Liu Yingjun, Li Fu-li and Zhu S Y 2013 Phys. Rev. A \textbf{87} 053831.

\bibitem {Qin:2014}Qin M, Shen H Z, Zhao X L and Yi X X 2014 Phys. Rev. E \textbf{90} 042140.

\bibitem {Zhang:2013}Zhang Yin-Ping, Li Hong-Rong, Fang Ai-Ping, Chen Hao and Li Fu-Li 2013 Chinese Physics B \textbf{22} 057104.

\bibitem {Wang:2013}Wang Xiao-Lian, Li Hong-Rong, Zhang Pei, Li Fu-Li 2013 Chinese Physics B \textbf{22} 7102.

\bibitem {Sarovar:2010}Sarovar M, Ishizaki A, Fleming G R and Whaley K B 2010 Nat. Phys. \textbf{6} 462.

\bibitem {Vos:1993}Vos M H, Rappaport F, Lambry J C, Breton J and Martin J L 1993 Nature. \textbf{363} 320.

\bibitem {Caruso:2010}Caruso F, Chin A W, Datta A, Huelga S F and Plenio M B 2010 Phys. Rev. A \textbf{81} 062346.

\bibitem {Ishizaki:2010}Ishizaki A and Fleming G R 2010 New J. Phys. \textbf{12} 055004.

\bibitem {Amerongen:2000}Amerongen H. van, Valkunas L, Grondelle R. van, 2000 Photosynthetic Excitons(Singapore:World Scientific).

\bibitem {Blankenship:2002}Blankenship R E 2002 Molecular mechanisms of photosynthesis(Wiley-Blackwell).


\bibitem {Wootters:1998}Wootters W K 1998 Phys. Rev. Lett. \textbf{80} 2245.


\end{thebibliography}
\end{document}